\documentstyle[12pt]{article}

\newcommand {\nn}    {\nonumber}
\newcommand {\vs}[1]  { \vspace*{#1 cm} }

\newcounter{eq}
\newcounter{sc}

\newcommand {\NP}   {Nucl.Phys.}
\newcommand {\PL}   {Phys.Lett.}

\def\overleftrightarrow#1{\vbox{\ialign{##\crcr
 $\leftrightarrow$\crcr\noalign{\kern-1pt\nointerlineskip}
 $\hfil\displaystyle{#1}\hfil$\crcr}}}

\newlength{\minitwocolumn}
\begin{document}


\begin{flushright}
EDO-EP-29\\
September, 1999\\
\end{flushright}
\vspace{30pt}

\pagestyle{empty}
\baselineskip15pt

\begin{center}
{\large\bf Mass Hierarchy and Trapping of Gravity

 \vskip 1mm
}

\vspace{20mm}

Ichiro Oda
          \footnote{
          E-mail address:\ ioda@edogawa-u.ac.jp. 
                  }
\\
\vspace{10mm}
          Edogawa University,
          474 Komaki, Nagareyama City, Chiba 270-0198, JAPAN \\

\end{center}


\vspace{15mm}
\begin{abstract}
We construct a model consisting of many D3-branes with only positive 
tension in a five-dimensional anti-de Sitter space-time geometry.
It is shown that this type of model naturally realizes not only 
exponential mass hierarchy between the Planck scale and the 
electroweak scale but also trapping of the graviton on the 
D3-branes. 
It is pointed out that our model may have a flexibility to 
explain the existence of more than one disparate mass scales,
such as the electroweak scale and the GUT scale, on the same
D3-brane.

\vspace{15mm}

\end{abstract}

\newpage
\pagestyle{plain}
\pagenumbering{arabic}


\rm
\section{Introduction}

Understanding the hierarchy problem of the Standard Model
in four dimensions is a classic and important problem.
The main ideas for this problem have been so far focused
on improvement of particle physics sector around the TeV scale.
Recently, however, motivated by studies of non-perturbative 
superstring theory and M-theory, a completely new proposal 
for the hierarchy problem has been made in gravity sector 
where the huge Planck scale is reduced all the way to the TeV scale 
by assuming the existence of extra dimensions 
\cite{Arkani, Antoni, Randall1}.

In particular, Randall and Sundrum have considered a solution for 
the five-dimensional Einstein equation with the cosmological constant 
and two 3-branes \cite{Randall1, Randall2}.
This solution involves a "red-shift" (or "warp") factor in the metric tensor, 
which was found to play a critical role in explaining the vast 
disparity between the Planck scale and the electroweak scale 
in a natural way \cite{Randall1}. Subsequently,
they proposed an alternative scenario for trapping of gravity
\cite{Randall2}. 

Here it is worth summarizing basic setups and aims 
of the two papers \cite{Randall1, Randall2}
since they have proposed two distinct new scenarios 
in terms of the same solution as stated above. 
In both the papers, two 3-branes, in other words, domain walls, are
placed at the boundaries of the fifth dimension, 
which are the fixed points of an orbifold $S^1/Z_2$, in the five-dimensional 
anti-de Sitter space-time \cite{Witten, Stelle}. 
The one 3-brane with positive brane tension
is located at the origin of the fifth dimension while the other
3-brane with opposite brane tension to the first
3-brane is located at another fixed point away from the 
origin at some distance. 
Note that one has to make such a fine-tuning of brane tensions
to obtain the Randall-Sundrum solution for the Einstein equation. 
(This issue will be argued in some detail around the end of section 2.) 

In the first paper \cite{Randall1}, our universe is assumed to the 3-brane
with negative tension whereas the other 3-brane with positive tension
is regarded as hidden universe. If we think that a natural scale on the
hidden world is of order the Planck scale, the electroweak scale is generated
on our universe from the red-shift factor by selecting rather small size of
an extra dimension.
Thus, the purpose of the paper is to present a resolution to the hierarchy
problem.

On the other hand, in the second paper \cite{Randall2}, the model setup is 
converse compared to that in the first paper \cite{Randall1}. 
That is, our universe is assumed to the 3-brane with positive 
tension at the origin whereas the other 3-brane with negative tension
is regarded as hidden universe. 
Moreover, the 3-brane with negative tension is moved to infinity,
thus providing an example of a non-compact extra dimension. A remarkable
thing here is that even without mass gap between the massless
graviton and the continuous Kaluza-Klein spectrum, the
four-dimensional Newton's law is reproduced to more than adequate
precision on our universe, thereby implying the trapping of gravity
on our universe. The reason why the 3-brane with positive tension is
taken to be our universe is that the 3-brane with positive tension
supplies us with the $\delta$-function potential with a negative
coefficient supporting the massless graviton, which will be discussed
in section 4. Of course, in this setup,
we cannot solve the hierarchy problem.

{}From these considerations, it is natural to ask whether one can construct
a model with the same geometry as in the Randall-Sundrum metric solution
such that problems of mass hierarchy and trapping of gravity on 
our 3-brane universe are simultaneously solved. 
Indeed, in a recent work \cite{Lykken},
Lykken and Randall have explored such a possibility by considering
a model including only 3-branes with positive tension. Their setup is
similar to that in the second paper \cite{Randall2}, but is distinct
in that the setup includes more than
one positive tension 3-brane. But it is unclear at least for the present
author that introduction of many 3-branes with positive tension in
addition to a single 3-brane with negative tension ("regulator brane")
in the setup is consistent with the Einstein equation.
One of motivations in this paper is to show explicitly that the above setup 
is indeed compatible with the Einstein equation. However, to do so, 
at the outset we need to introduce the same number of negative tension
3-branes as positive tension 3-branes into a model, and after that we
have to take a suitable limit in order to move negative tension 3-branes
to infinity. 
Of course, in an extreme case where all the negative tension 3-branes are 
coincident, our model would become equivalent to that of Lykken and Randall.

The other motivations behind the study at hand are as follows.
In this paper, we make use of solutions satisfying the Einstein equation,
which were found in our previous work \cite{Oda1}. As stated in the
paper, these new solutions describe many domain walls standing along
the fifth dimension with topology $S^1$ in five-dimensional anti-de
Sitter space-time, so they realize many universe cosmology. But these
solutions involve the same number of positive tension 3-branes as
negative tension 3-branes, thus as a result, the total number of 3-branes
are only even. It is quite unfair  and against democracy
that only even universes can exist in such a model. 
It will be shown later that
we can construct a more plausible model with any number of
D3-branes with positive tension by moving negative tension O3-planes
to infinity. 

Another important motivation of this study is relevant to a recent
interesting work by Dienes et al \cite{Dienes}. In their paper, some 
phenomenological difficulties associated with the Randall-Sundrum scenario 
and its extension
were pointed out and possible resolutions to these puzzles are speculated.
Specially, they have concluded that in the Randall-Sundrum model we
cannot simultaneously generate the Planck/electroweak hierarchy 
(mass hierarchy) $\it{and}$ explain gauge coupling unification. This
conclusion is physically sensible, of course, 
as in the original Randall-Sundrum
model, there exists only a single red-shift factor depending on a relative
distance between the two 3-branes. 
This red-shift factor operates universally on all mass scales in
our universe so it is difficult to generate two mass scales, such as
the electroweak scale and the GUT scale, on our universe
without recourse to additional mechanisms. 
One interesting resolution to this puzzle is to consider
at least three universe model where one D3-brane is regarded as our 
universe and the remaining two D3-branes are taken to be hidden
universes. As there are two red-shift factors in this model, one
could explain the Planck/electroweak hierarchy as well as the
existence of the GUT scale on our universe at the same time \cite{Oda2}. 

The organization of the paper is as follows: in section 2 we
review our previous study \cite{Oda1}. Here new solutions
for the Einstein equation are presented. Moreover, it is
stated in detail about the reason why it is difficult to construct
a model with only positive tension branes. In section 3, based on
the solutions reveiwed in section 2, we construct a model with
two positive tension 3-branes in a concrete way. The generalization
to arbitrary number of positive tension branes is also discussed. 
In section 4,
we investigate mass hierarchy and trapping of gravity along a
similar line of arguments to the previous works
\cite{Randall2, Lykken}.  The final
section is devoted to discussions and future directions of this 
work.

\section{Review of many domain wall model}

We begin by briefly reviewing a model of Ref.\cite{Oda1}.
This will enable us to establish our notations and conventions
and explain why it is difficult to construct a model
consisting of only D3-branes with positive tension.

Our starting action is the Einstein-Hilbert action with 
the cosmological constant in five dimensions plus an action 
describing many domain walls in four dimensions \cite{Oda1}:
\begin{eqnarray}
S = \frac{1}{2 \kappa^2} \int d^4 x \int_{0}^{2L} dz 
\sqrt{-g} \left(R - 2 \Lambda \right) 
+ \sum_{i=1}^{n} \int_{z=L_i} d^4 x \sqrt{-g_i} {\cal L}_i,
\label{1}
\end{eqnarray}
where the cosmological constant $\Lambda$ is taken to a negative
number,
which implies that the geometry of five-dimensional bulk is 
anti-de Sitter space-time. 
The fifth dimension $z$ is assumed to be compact with the length $2L$,
but later regarded to be effectively non-compact by taking the limit
$L \rightarrow \infty$.
Moreover, $\kappa$ denotes the five-dimensional gravitational
constant with a relation $\kappa^2 = 8 \pi G_N = \frac{8 \pi}{M_*^3}$ 
where $G_N$ and $M_*$ are the five-dimensional Newton constant
and the five-dimensional Planck scale, respectively. 
Throughout this article we follow the standard 
conventions and notations of the textbook of Misner, Thorne and 
Wheeler \cite{Misner}. 
Note one important distinction between our model (\ref{1}) and
the Randall and Sundrum model \cite{Randall1, Randall2}. 
In the Randall and Sundrum model \cite{Randall1, Randall2} the geometry
of the fifth dimension is taken to a singular orbifold geometry
$S^1/Z_2$ \cite{Witten, Stelle}, whereas in our model the topology
of the fifth dimension is a circle $S^1$ because the existence
of many domain wall solution for the Einstein equation requires
us to choose this smooth manifold.

Variation of the action (\ref{1}) with respect to the five-dimensional
metric tensor leads to the Einstein equation:
\begin{eqnarray}
\sqrt{-g} \left( R^{MN} - \frac{1}{2} g^{MN} R \right)
= - \sqrt{-g} g^{MN} \Lambda 
+ \kappa^2 \sum_{i=1}^{n}  \sqrt{-g_i} g_i^{\mu\nu}
\delta_{\mu}^M \delta_{\nu}^N {\cal L}_i \delta(z - L_i),
\label{2}
\end{eqnarray}
where $M, N, ...$ denote five-dimensional indices, whereas
$\mu, \nu, ...$ do four-dimensional ones. 
Provided that we adopt a metric ansatz
\begin{eqnarray}
ds^2 &=& g_{MN} dx^M dx^N  \nn\\
&=& u(z)^2 \eta_{\mu\nu} dx^\mu dx^\nu + dz^2,
\label{3}
\end{eqnarray}
with $\eta_{\mu\nu}$ denoting the four-dimensional Minkowski metric,
the Einstein equation (\ref{2}) reduces to two combined differential 
equations for the unknown function $u(z)$:
\begin{eqnarray}
\left(\frac{u'}{u} \right)^2  = k^2,
\label{4}
\end{eqnarray}
\begin{eqnarray}
\frac{u''}{u} 
= k^2 + \frac{\kappa^2}{3} \sum_{i=1}^{n} {\cal L}_i \delta(z - L_i),
\label{5}
\end{eqnarray}
where the prime denotes a differentiation with respect to $z$ 
and we have defined $k$ as
\begin{eqnarray}
k = \sqrt{- \frac{\Lambda}{6}}.
\label{6}
\end{eqnarray}

In the previous paper \cite{Oda1}, we have seeked special solutions 
with the form of
\begin{eqnarray}
u(z) =  e^{- k  f(z)},
\label{7}
\end{eqnarray}
and we have presented two distinct solutions with simple and 
manageable form although the other complicated solutions could
be also constructed.
These solutions describe even domain walls standing along $S^1$ at 
some intervals in five-dimensional anti-de Sitter space-time.
One solution describes $\frac{n-1}{2}$ even domain walls locating 
at $L_{2i}$ and is concretely given by
\begin{eqnarray}
f(z) &=& |z| +  \sum_{i=1}^{\frac{n-1}{2}} (-1)^i |z - L_{2i}| - L, 
\nn\\
f'(z) &=& \sum_{i=1}^{\frac{n-1}{2}} (-1)^i \varepsilon(z - L_{2i}) + 1, 
\nn\\
f''(z) &=& 2 \sum_{i=1}^{\frac{n-1}{2}} (-1)^i \delta(z - L_{2i}),
\label{8}
\end{eqnarray}
for which ${\cal L}_i \
(i = 1, 2, \cdots, \frac{n-1}{2})$ must satisfy the relations
\begin{eqnarray}
{\cal L}_{2i} = (-1)^{i+1} \frac{6k}{\kappa^2}.
\label{9}
\end{eqnarray}
The other solution describes $n-1$ even domain walls 
locating at $L_i \ (i=1, 2, \cdots, n-1)$ and
takes the form
\begin{eqnarray}
f(z) &=& \sum_{i=2}^{n-1} (-1)^{i+1} |z - L_i| + L, 
\nn\\
f'(z) &=& \sum_{i=1}^{n-1} (-1)^{i+1} \varepsilon(z - L_i) - 1, 
\nn\\
f''(z) &=& 2 \sum_{i=1}^{n-1} (-1)^{i+1} \delta(z - L_i),
\label{10}
\end{eqnarray}
for which, this time, ${\cal L}_i \ (i = 1, 2, \cdots, \frac{n-1}{2})$ 
must satisfy the relations
\begin{eqnarray}
{\cal L}_{2i} = - {\cal L}_{2i-1} = \frac{6k}{\kappa^2}.
\label{11}
\end{eqnarray}
Moreover, in the both solutions, $L_i$ satisfies the relations
\begin{eqnarray}
L_{2i} = \frac{L_{2i-1} + L_{2i+1}}{2}, \ L_1 \equiv 0, 
\ L_n \equiv 2L,
\label{12}
\end{eqnarray}
with $i = 1, 2, \cdots, \frac{n-1}{2}$. 

To close this section, it is worthwhile to argue why it is 
difficult to construct a model consisting of only D3-branes with 
positive tension. This problem is indeed closely related to
various important problems associated with the models under 
consideration.
Before doing that, let us recall that
in the original Randall-Sundrum model \cite{Randall1,Randall2}, 
there are two 3-branes with opposite sign of
brane tension at the boundaries of an orbifold $S^1/Z_2$.
Also in our model, there are the same number of 3-branes
with positive tension and negative tension on a circle
$S^1$. It was stressed in Ref.\cite{Lykken} that 
the necessity for the negative energy objects might be
one of disadvantages in those setups since not only it is 
believed that the Standard Model is placed on the D3-brane, 
which is certainly a positive energy object, but also
some problematic facts were pointed out in the cosmological context.
For instance, the Friedmann-like expanding 
universe does not arise if the Standard Model is placed 
on the 3-brane with negative brane tension \cite{Csaki, Grojean}. 
Actually, in Ref.\cite{Randall1}, the Standard Model was
located on such a negative energy 3-brane in order to
explain the exponential mass hierachy. 

Keeping these facts in mind, let us now turn to our question:
"Why is it difficult to construct a model consisting of only
many D3-branes with positive tension?"  The answer can be 
easily found from considerations of the electro-maganetics 
in a compact space. It is well known that we cannot put a 
single point
charge in a $\it{compact}$ space since electric flux lines have no
place to go in the compact space. In other words, the field
equation does not admit the existence of such a configuration.
To remedy it, the simplest way is to introduce another 
point charge with
the same size but opposite sign to the first point charge in order
to make the flux lines exactly close. The other simplest way is
to take account of a non-compact space by taking the limit of infinitely
large size. In a non-compact space, the flux lines arising from
a single point charge could run to infinity so we could have a consistent
model with any number and configuration of 3-branes. In fact, this latter 
procedure has been recently taken by Lykken and Randall \cite{Lykken} 
at least implicitly in order to make a model with
only positive tension 3-branes. 

Incidentally, it is of interest to observe that a similar situation 
also occurs in the context of D-brane theory \cite{Polchinski}.   
The D-brane, unlike the fundamental string, carries positive R-R charge.
We cannot therefore put a single (or many) D-brane(s) in a compact 
region owing to a non-zero total R-R charge. To cancel the R-R charge 
exactly, we are led to
introduce some objects which carry negative R-R charge with the same size.
They are nothing but orientifolds!  From this analogy, it is interesting
to regard 3-branes with positive and negative tension as D3-branes and
O3-planes, respectively. We think that we should pursue this analogy
further to understand a long-standing problem, i.e., the cosmological 
constant problem, in future.

\section{A model with positive tension D3-branes}

In this section, on the basis of the solutions given in the previous
section, we will present a concrete model which realizes simultaneously 
mass hierarchy and trapping of gravity on our universe 
("visible 3-brane"). An essential aspect with respect to the two
phenomena is very similar in the two distinct solutions 
Eqs.(\ref{8}), (\ref{10}), so we limit ourselves to the type of
the solution Eq.(\ref{10}) in this paper. Moreover, in this section
and the next section, to start with, 
we shall give a model in the case of two positive
tension 3-branes and then extend it to a more general case of
arbitrary number of positive tension 3-branes.

Let us start with a model with two positive tension branes 
and two negative tension branes, which is a specific example
($n=5$) of the solution (\ref{10}):
\begin{eqnarray}
f(z) &=& - |z - L_2| + |z - L_3| - |z - L_4| + L,  \nn\\
f'(z) &=& \varepsilon(z) - \varepsilon(z - L_2) + \varepsilon(z - L_3)
- \varepsilon(z - L_4) - 1, \nn\\
f''(z) &=& 2 \left[ \delta(z) - \delta(z - L_2) + \delta(z - L_3)
- \delta(z - L_4) \right].
\label{13}
\end{eqnarray}
Of course, ${\cal L}_i$ must satisfy the relation (\ref{11})
specified to $n=5$ case:
\begin{eqnarray}
{\cal L}_1 = - {\cal L}_2 = {\cal L}_3 = - {\cal L}_4 
= - \frac{6k}{\kappa^2}.
\label{14}
\end{eqnarray}
Here, instead of Eq.(\ref{12}) we require slightly modified 
relations for $L_i$: 
\begin{eqnarray}
L = L_2 - L_3 + L_4, \ L_1 \equiv 0, \ L_n \equiv 2L, \
L_3 > 2 L_2.
\label{15}
\end{eqnarray}
The reason is that as mentioned in the paper \cite{Oda1},
the solution (\ref{10}) with the relation (\ref{12}) has
a characteristic feature $f(L_{2i-1}) = 0$ (or equivalently, 
$u(L_{2i-1}) = 1$)
which is a undesired feature in explaining mass hierarchy.
Thus, in order to avoid this feature, we have imposed the relation
(\ref{15}), specially, $L_3 > 2 L_2$ on the solution (\ref{13}). 
Indeed, as shown in the next section,
this type solution has a desired feature
with respect to both mass hierarchy and trapping of gravity.
And note that the positive tension domain walls are located at
$z = L_1 \equiv 0, L_3$ while the negative tension domain walls
are at $z = L_2, L_4$ as seen in Eq.(\ref{14}). For comparison
with Ref.\cite{Lykken}, according to their terminology, we
may call the positive tension 3-branes at $z = L_1 \equiv 0$ and
at $z = L_3$ "TeV brane" and "Planck brane", respectively 
\footnote{Our setup is different from that of Ref.\cite{Lykken}
where "Planck brane" and "TeV brane" are put at $z = L_1 \equiv 0$ 
and at $z = L_3$, respectively. One can modify the present setup
to coincide with their setup without any difficulty.}.  
It might appear to be difficult to move only the two negative
tension branes to infinity without changing the essential
contents of the model since the negaive and the positive tension
branes are adjacent in the solution (\ref{13}). 
But the analysis in the next section
reveals that this procedure can be carried out by taking
the limit of $L_2, 2L - L_4 \rightarrow \infty$ with keeping
$L_3 - 2 L_2$ finite. In this way, we can construct a model
consisting of only two positive tension domain walls separated
at some distance along a non-compact fifth dimension.

Next let us present a model including general $\frac{n-1}{2}$ 
positive tension domain walls which is a straightforward 
generalization of the model with two positive tension domain 
walls.
This model is just given in terms of Eq.(\ref{10}) satisfying
Eq.(\ref{11}) but modified relations among $L_i$ 
compared to Eq.(\ref{12}), and a limiting procedure. 
For convenience, let us write down explicitly this 
general model with $\frac{n-1}{2}$ positive tension D3-branes:
\begin{eqnarray}
f(z) &=& \sum_{i=2}^{n-1} (-1)^{i+1} |z - L_i| + L, 
\nn\\
f'(z) &=& \sum_{i=1}^{n-1} (-1)^{i+1} \varepsilon(z - L_i) - 1, 
\nn\\
f''(z) &=& 2 \sum_{i=1}^{n-1} (-1)^{i+1} \delta(z - L_i),
\nn\\
{\cal L}_{2i} &=& - {\cal L}_{2i-1} = \frac{6k}{\kappa^2},
\nn\\
L &=& \sum_{i=2}^{n-1} (-1)^i L_i, \ L_1 \equiv 0, 
\ L_n \equiv 2L,
\label{13-2}
\end{eqnarray}
with $i = 1, 2, \cdots, \frac{n-1}{2}$. And the limiting procedure
is given by
\begin{eqnarray}
f(L_{2i}) \rightarrow \infty,
\label{13-3}
\end{eqnarray}
with keeping $f(L_{2i-1})$ some finite, negative values.

\section{Exponential mass hierarchy and Newton's law}

In the previous section, we have presented a concrete model
so we are now ready to consider how this model resolves 
problems of mass hierarchy and trapping of gravity.
As in the previous section, let us start with the model with
two positive tension domain walls. Following the formula
given in Ref.\cite{Oda1}, it is straightforward to evaluate
mass scale $m(0)$ on the "TeV brane" located at $z=L_1 \equiv 0$
from mass scale $m(L_3)$ on the "Planck brane" located at $z=L_3$
to which the Planck mass scale is allocated. The result is
of the form
\begin{eqnarray}
m(0) = e^{-k(L_3 - 2 L_2)} m(L_3).
\label{16}
\end{eqnarray}
As stated in the previous section, since we have taken $L_3 - 2 L_2$
to be a positive and finite value, we can resolve the mass hierarchy 
if $L_3 - 2 L_2$ is of order of $10$. 
In other words, the mass scale in our universe
("TeV brane") is of order electroweak scale thanks to the red-shift
factor in the geometry when the Planck mass scale is assigned to 
another positive tension domain wall ("Planck brane")
at $z = L_3$ and $L_3 - 2 L_2$ is of order of $10$.

Next, let us turn our attention to a problem of trapping of gravity
on "TeV brane".
To so that, we will consider the linearlized approximation of the 
metric tensor and examine small fluctuations $h_{\mu\nu}$ around
the four-dimensional Minkowski metric on the brane and determine
the graviton spectrum as well as the Kaluza-Klein spectrum.
Accordingly, we assume the form 
\begin{eqnarray}
ds^2 &=& g_{MN} dx^M dx^N  \nn\\
&=& e^{-2k f(z)} \left( \eta_{\mu\nu} + h_{\mu\nu}(x, z) \right)
dx^\mu dx^\nu  + dz^2.
\label{17}
\end{eqnarray}
Then, with the gauge conditions $h_\mu^\mu = \partial^\mu h_{\mu\nu}
= 0$, up to the leading order of $h_{\mu\nu}$, the field equation is
of the form
\begin{eqnarray}
-\frac{1}{2} e^{2k f(z)} \Box h_{\mu\nu} - \frac{1}{2} h_{\mu\nu}''
+ 2k^2 h_{\mu\nu} - k f''(z) h_{\mu\nu} = 0, 
\label{18}
\end{eqnarray}
where $\Box$ denotes the flat space-time four-dimensional Laplacian 
operator, and $f(z)$ is defined as in Eq.(\ref{13}).
At this stage, if we consider the plane wave fluctuations
\begin{eqnarray}
h_{\mu\nu}(x, z) =  e^{i p_\mu x^\mu} h_{\mu\nu}(z), \
p_\mu^2 = - m^2, 
\label{19}
\end{eqnarray}
Eq.(\ref{18}) reduces to the differential equation for 
$h_{\mu\nu}(z)$
\begin{eqnarray}
\left[-\frac{m^2}{2} e^{2k f(z)} - \frac{1}{2} \partial_z^2
+ 2k^2  - k f''(z) \right] h_{\mu\nu}(z) = 0, 
\label{20}
\end{eqnarray}

It is more convenient to rewrite Eq.(\ref{20}) into the one-dimensional
Schrodinger wave-equation by making a change of variables,
$y = \frac{1}{k} e^{k f(z)}$
\begin{eqnarray}
\left[-\frac{1}{2} \partial_y^2 + V(y) \right] \Psi(y) = 
\frac{m^2}{2} \Psi(y), 
\label{21}
\end{eqnarray}
where we have defined as 
\begin{eqnarray}
h_{\mu\nu}(z) \equiv k^{-\frac{1}{2}}
y^{-\frac{1}{2}} \Psi_{\mu\nu}(y), \
\Psi_{\mu\nu}(y) \equiv
\Psi(y).
\label{22}
\end{eqnarray}
 The one-dimensional potential is now given by
\begin{eqnarray}
V(y) = \frac{15}{8} \frac{1}{y^2} - \frac{2}{y} \left[
\delta(y - y_1) - \delta(y - y_2) + \delta(y - y_3) - 
\delta(y - y_4) \right], 
\label{23}
\end{eqnarray}
where $y_i (i=1,2,3,4)$ are of the form 
\begin{eqnarray}
y_1 &=& \frac{1}{k}, \nn\\
y_2 &=& \frac{1}{k} e^{k L_2}, \nn\\
y_3 &=& \frac{1}{k} e^{-k (L_3 - 2L_2)}, \nn\\
y_4 &=& \frac{1}{k} e^{k (2L - L_4)}, 
\label{24}
\end{eqnarray}
thus we have a relation
\begin{eqnarray}
0 < y_3 < y_1 < y_2 < y_4,
\label{25}
\end{eqnarray}
where without loss of generality we have assumed $y_2 < y_4$. 
Here an interesting thing has happened
owing to Eq.(\ref{15}).
Namely, the change of variables from $z$ to $y$ has caused
a rearrangement of the positions of domain walls where the two
positive tension 3-branes are located at the left side of
the two negative tension 3-branes, so we can move the two negative 
tension 3-branes to infinity by taking the limit of 
$L_2, 2L - L_4 \rightarrow \infty$ with keeping $L_3 - 2 L_2$ finite. 
Note that the two positive 
tension 3-branes stay at the same points in this limit.

The one-dimensional Schrodinger wave-equation Eq.(\ref{21}) gives
rise to much useful informations about the graviton and a tower
of the Kaluza-Klein modes. Luckily enough, the potential $V(y)$ has
a similar form to that in Refs.\cite{Randall2, Lykken} so
that we can take over the results obtained there to the present
case with an appropriate modification.
First of all, the potential $V(y)$ has the $\delta$-functions
at $y = y_1, y_3$ with negative coefficients, which means
that these $\delta$-functions supports a normalizable bound
state mode, which is of course nothing but the massless graviton.
Incidentally, the "regulator" branes with negative brane tension
locating at $y = y_2, y_4$, though they are moved to infinity,
induce the $\delta$-functions with positive coefficients into
the potential $V(y)$, so these branes cannot support such a
massless graviton. This is the reason that in Ref.\cite{Randall2},
Randall and Sundrum have regarded the 3-brane with positive 
tension as our universe for the trapping of graviton on the
brane.

Second, let us focus our attention to the properties of 
the continuous KK modes.
In this point, in the leading approximation, we can safely 
neglect the existence of the $\delta$-functions in the 
potential $V(y)$. Then, it turns out that the squared mass
$m^2$ of the KK modes is positive definite as desired.
In the non-compact
limit, the potential falls off to zero, so there is no mass gap
between the massless graviton and the KK modes. At first sight,
this could be a signal of danger since a tower of KK modes 
would give measurable effects to the modification 
of the Newton's law, but as shown shortly, 
they give rise to only small corrections to
the Newton's law \cite{Randall2, Lykken}. We can easily write
down the general solution for the continuous KK modes as
\begin{eqnarray}
\Psi_m(y) \sim \frac{m^{\frac{5}{2}}}{k^2} \sqrt{y}
\left[Y_2(my) + \frac{4k^2}{\pi m^2} J_2(my) \right],
\label{26}
\end{eqnarray}
where $Y_2$ and $J_2$ denote the Bessel functions of order 2.
{}From the above wave function, it is straightforward to
evaluate the corrections to the Newton's law from the continuous
KK modes \cite{Randall2, Lykken}. In fact, the gravitational
potential between two masses $m_1$ and $m_2$ takes the form
in the polar coordinate
\begin{eqnarray}
U(r) = G_N \frac{m_1 m_2}{r} + \int_0^\infty \frac{dm}{k} G_N
\frac{m_1 m_2 e^{-mr}}{r} = G_N \frac{m_1 m_2}{r} \left(1 + 
\frac{1}{k^2 r^2} \right).
\label{27}
\end{eqnarray}
Accordingly, it turns out that an observer living in the 
"TeV brane" (and also "Planck brane") sees gravity as essentially
four-dimensional since the corrections to the Newton's law 
from the KK modes are very small. (Note that $k$ is taken to
be of order the Planck scale.)
 
In the above, the case of the two D3-branes with positive 
tension has been investigated in detail. 
The analysis of the general model Eqs.(\ref{13-2}) and (\ref{13-3}) 
is quite straightforward and very similar to the case of
the two D3-branes, so that we will not repeat it in this paper.
For instance, in the Schrodinger wave-equation (\ref{21}),
only the modification lies in the form of the potential
$V(y)$ where $n-1$ $\delta$-functions appear instead
of 4 $\delta$-functions. Hence, we can show the trapping of gravity
on the D3-branes again. 
On the other hand, for the hierarchy problem, one can also
show that the electroweak scale is generated on our universe
when the Planck scale is assigned to a hidden universe and
an appropriate value of distance between the two universes
is chosen.
However, in this general model, this is not the whole story
since there are many red-shift factors arising from
relative distances between our universe and the remaining
hidden universes.
This issue is worthy of further study and will be reported
in detail in a separate publication \cite{Oda2}.

\section{Discussions}

In this paper, we have investigated  problems of both mass hierarchy
and trapping of gravity on our universe by using the solutions
describing many domain walls which were previously found by the present
author \cite{Oda1}.
Our model in general consists of any number of positive tension
D3-branes in a non-compact extra dimension, and shows the exponential
mass hierarchy and trapping of gravity in our universe.
In this sense,
the study at hand shows that the scenario by Lykken and
Randall \cite{Lykken} is indeed realized.

Since various matters and gauge fields are also localized
on our D3-brane universe in terms of the mechanism in string
theory, our model is equipped with necessary conditions as
a realistic model. Thus, future directions of this work would
be to apply our model to other unsolved problems.
In fact, we can easily point out that our model has the 
following advantages over the original Randall-Sundrum model.
In the context of cosmology, since there exist only positive 
tension D3-branes in our model,
the lack of the Friedmann-like expanding universe is certainly
resolved \cite{Csaki, Grojean}.
And, in the context of phenomenology, some of puzzles stated in
a recent study \cite{Dienes} seem to be also resolved in terms of 
our general model including many D3-branes since we have not a
single but many "red-shift" factors associated with many
domain walls. These problems will be reported in a separate 
publication \cite{Oda2}.

\vs 1

\end{document}